\documentclass[seceq]{ptptex}

\title{
An Analytic Analysis of Phase Transitions \\ in Holographic Superconductors
}

\author{
Chiang-Mei \textsc{Chen}$^{1,2,}$\footnote{E-mail: cmchen@phy.ncu.edu.tw} and
Ming-Fan \textsc{Wu}$^{1,}$\footnote{E-mail: 93222036@cc.ncu.edu.tw}
}

\inst{
$^1$Department of Physics, National Central University, Chungli 320, Taiwan \\
$^2$Center for Mathematics and Theoretical Physics, National Central University, Chungli 320, Taiwan
}

\abst{
Using a simple analytic approach, we study the universal properties of second-order phase transition in holographic superconductor models. We explore a general model in arbitrary dimensions in which the condensation occurs via the St\"uckelberg spontaneous symmetry breaking mechanism. All the possible second-order phase transitions and their universal characteristics can be identified analytically. The relationship between the critical temperature and charge density is generic, and the critical exponents can be greater than the typical mean field value $1/2$. In addition, the related numerical factors can also be computed qualitatively.
}

\begin{document}

\maketitle

\section{Introduction}
In the past years, the anti-de Sitter/conformal field theory (AdS/CFT) correspondence~\cite{Maldacena:1997re, Gubser:1998bc, Witten:1998qj} was intensively applied to study strongly coupled phenomena in various physical problems, such as in quantum chromodynamics (QCD), condensed matter physics etc. In virtue of the remarkable strong/weak duality, the AdS/CFT provides an astonishing advance to deal with certain field theories in the strong coupling region via their holographic dual weakly coupled gravitational descriptions. In particular, the investigation of the phase transitions in holographic superconductors has achieved significant progress, see excellent reviews~\cite{Hartnoll:2009sz, Herzog:2009xv, Horowitz:2010gk}. The pioneering scheme of the gravitational dual description~\cite{Gubser:2008px, Hartnoll:2008vx}, the so-called abelian Higgs model, considered a scalar field and a U(1) gauge field in the planar Schwarzschild-anti-de Sitter black hole. A nontrivial profile of the scalar field manifests the condensate of the paring mechanism, and the U(1) gauge field exhibits electromagnetic properties in superconductors.

One of the most interesting phenomena in holographic superconductor research is the second-order phase transition. The correlation lengths near the critical point of second-order phase transitions become divergent, so the systems can be described by a scale-invariant theory. In such circumstances, the AdS/CFT is an exceptionally appropriate technique to investigate the corresponding universal natures, in particular, the critical exponents. It turns out that the critical exponent in the abelian Higgs model is $1/2$, which is the universal value in the mean field theory. Later, a generalized model~\cite{Franco:2009yz, Franco:2009if} was proposed in which the condensation occurs via the St\"uckelberg spontaneous symmetry breaking mechanism. The new features of the St\"uckelberg holographic superconductor models include the presence of first-order phase transitions and second-order phase transitions with nonmean field behavior~\cite{Franco:2009yz, Franco:2009if}.

Moreover, some analytic approaches have been proposed to address the universal properties of second-order phase transitions in holographic superconductors. One suggestion is to consider a simple match of the boundary and horizon solutions of both the scalar and U(1) gauge fields~\cite{Gregory:2009fj}, see also Refs.~\cite{Ge:2010aa, Bellon:2010xs}. Another analytic approach introduces a trial function with a free parameter which that is suggested to be fixed by a minimizing procedure~\cite{Siopsis:2010uq}, see also Refs.~\cite{Zeng:2010zn, Li:2011xj}. In this work, we apply the simple matching approach to analytically study the second-order phase transition in a general class of the St\"uckelberg holographic superconductor in arbitrary dimensions. We examine all the possibilities that the second-order can occur, and derive the corresponding critical exponents. Our results are consistent with the numerical analysis~\cite{Franco:2009yz, Franco:2009if}. Some related properties in three- and four-dimensional spacetimes have been discussed in~\cite{Herzog:2010vz, Aprile:2010yb}.

The outline of the paper is as follows. In $\S2$, we review the St\"uckelberg holographic superconductor models. The analytic study of phase transition is given in $\S3$, and explicit models are examined in $\S4$. Some discussions are given in the last section.

\section{St\"uckelberg holographic superconductor}
The St\"uckelberg holographic superconductor~\cite{Franco:2009yz, Franco:2009if} is a simple generalization of the abelian Higgs model~\cite{Gubser:2008px, Hartnoll:2008vx, Horowitz:2010gk} in which the condensation occurs via the St\"uckelberg spontaneous symmetry breaking mechanism. In the probe limit, the action of the model can be decoupled into two parts. The gravitational section provides the background of a black hole in the anti-de Sitter (AdS) space providing the temperature in the holographic superconductor. The action of this section, in general (d+1)-dimensions, is
\begin{equation}
S_g = \int d^{d+1}x \sqrt{-g} \left( R - 2 \Lambda \right),
\end{equation}
and the gravitational background is a planar Schwarzschild-anti de Sitter black hole
\begin{equation} \label{SAdS}
ds^2 = - f(r) dt^2 + f^{-1}(r) dr^2 + r^2 \left( dx_1^2 + \cdots + dx_{d-1}^2 \right),
\end{equation}
where (the radius of AdS is fixed to be unity, i.e., $L^2 = - d (d - 1)/2\Lambda = 1$)
\begin{equation}
f(r) = r^2 \left( 1 - \frac{r_H^d}{r^d} \right).
\end{equation}
The parameter $r_H$ represents the radius of the black hole, which is determined by its mass $r_H^d = M^{d-2}$. The Hawking temperature of the black hole is
\begin{equation}
T = \frac{d}{4 \pi} r_H.
\end{equation}
In the later analysis, it is more convenient to redefine the radial coordinate as
\begin{equation}
z = \frac{r_H}{r},
\end{equation}
such that the locations of the boundary (asymptotic) and horizon are at $z = 0$ and $z = 1$, respectively.

The second part generically includes a pair of real scalar fields, $(\psi, \varphi)$, and one U(1) gauge field, $A_\mu$, in response to the condensate and conductivity in the superconductor. All the fields are assumed as probes and their back reaction to gravity can be neglected. The action of this section is
\begin{equation}
S_m = \int d^{d+1}x \sqrt{-g} \left( -\frac14 F^2 - \frac12 (\partial\psi)^2 - \frac{m^2}2 \psi^2 - \frac12 {\cal F}(\psi) (\partial \varphi - A)^2 \right).
\end{equation}
In this model~\cite{Franco:2009yz}, the dynamics is significantly dependent on the choice of the function ${\cal F}(\psi)$. The abelian Higgs model indeed corresponds to a particular choice of ${\cal F}(\psi) = \psi^2$. In this paper, we will focus only on the case that ${\cal F}(\psi)$ is a polynomial of $\psi$. The corresponding equations of motion are
\begin{eqnarray}
\frac1{\sqrt{-g}} \partial_\mu \left( \sqrt{-g} F^{\mu\nu} \right) + {\cal F}(\psi) (\partial^\nu \varphi - A^\nu) &=& 0,
\\
\partial_\mu \left[ \sqrt{-g} {\cal F}(\psi) (\partial^\mu \varphi - A^\mu) \right] &=& 0,
\\
\frac1{\sqrt{-g}} \partial_\mu \left( \sqrt{-g} \partial^\mu \psi \right) - m^2 \psi - \frac12 {\cal F}'(\psi) (\partial \varphi - A)^2 &=& 0.
\end{eqnarray}

By the gauge freedom, one can assume $\varphi = 0$. Moreover, we will only consider the scalar potential excitation of the U(1) gauge field, i.e., $A = \phi \, dt$. Therefore, the equations of motion in the gravitational background of~(\ref{SAdS}) reduce to, in terms of the coordinate $z$,
\begin{eqnarray}
\partial_z^2 \phi - \frac{d-3}{z} \partial_z \phi - \frac{{\cal F}(\psi)}{z^2 (1 - z^d)} \phi &=& 0,
\\
\partial_z^2 \psi - \frac{d - 1 + z^d}{z (1 - z^d)} \partial_z \psi - \frac{m^2}{z^2 (1 - z^d)} \psi + \frac12 \frac{{\cal F}'(\psi) \phi^2}{r_H^2 (1 - z^d)^2} &=& 0.
\end{eqnarray}

\section{Analytic study of phase transition}
A simple analytic approach by taking a trivial match of asymptotic and near-horizon solutions has been applied to address the properties of the phase transition in holographic superconductors~\cite{Gregory:2009fj}. It turns out that this simple analysis can really capture the universal properties of second-order phase transitions. In this section, we are going to reproduce such study for the St\"uckelberg holographic superconductor model.

At the boundary, $z = 0$, the solutions of $\phi$ and $\psi$ have the following asymptotic forms
\begin{eqnarray}
\psi_B(z) &=& D_+ z^{\lambda_+} + D_- z^{\lambda_-},
\\
\phi_B(z) &=& \mu - \frac{\rho}{r_H^{d-2}} z^{d-2},
\end{eqnarray}
where
\begin{equation} \label{lambda}
\lambda_\pm = \frac{d \pm \sqrt{d^2 + 4 m^2}}{2}.
\end{equation}
The parameters $\mu$ and $\rho$ are interpreted as the chemical potential and the charge density of the dual theory on the boundary. There are two possible condensates of the scalar field $\psi$ by turning on either the $D_+$ or $D_-$ mode. Then, the associated vacuum expectation values of the dual operators are
\begin{equation}
\langle {\cal O}_\pm \rangle = r_H^{\lambda_\pm} D_\pm.
\end{equation}
The parameters $\lambda_\pm$ define the dimensions of operators ${\cal O}_\pm$. From the expression for $\lambda_\pm$, the mass of the scalar $\psi$ should be $m^2 > -d^2/4$, which is consistent with the Breitenlohner-Freedman (BF) bound by stability of a scalar field in AdS$_{d+1}$.
A typical choice in the literature is $m^2 = -2$.

At the horizon, $z = 1$, the solutions can be expanded as
\begin{eqnarray}
\phi_H(z) &=& \phi_H(1) - \phi_H'(1) (1 - z) + \frac12 \phi_H''(z) (1 - z)^2 + \cdots,
\\
\psi_H(z) &=& \psi_H(1) - \psi_H'(1) (1 - z) + \frac12 \psi_H''(z) (1 - z)^2 + \cdots.
\end{eqnarray}
To be regular at the horizon, we should impose an additional condition $\phi_H(1) = 0$. After that, by solving the equations of motion order by order of $1 - z$, the expanding coefficients are related as, denoting $a = \phi_H'(1)$ and $b = \psi_H(1)$,
\begin{eqnarray}
\phi_H''(1) &=& (d - 3) a - \frac1{d} a {\cal F}(b),
\\
\psi_H'(1) &=& - \frac{m^2}{d} b,
\\
\psi_H''(1) &=& \frac{m^2 (m^2 + 2 d)}{2 d^2} b - \frac1{4 d^2 r_H^2} a^2 {\cal F}'(b).
\end{eqnarray}

The essential step in any analytic study is to construct a suitable scheme to resolve the relations of the coefficients in the boundary and horizon solutions. For our simple approach, we just take a trivial manner to match the boundary and horizon solutions of $\phi$ and $\psi$, and their first derivatives at a given matching point $z_m$ in the range of $0 < z_m < 1$. It turns out that different choices of the matching point will give different numerical coefficients, but the universal properties, such as the critical exponent, are actually independent of the choice of $z_m$. Let us consider a generic choice of $z_m$, and the matching conditions
\begin{equation}
\phi_B(z_m) = \phi_H(z_m), \quad \phi_B'(z_m) = \phi_H'(z_m); \quad \psi_B(z_m) = \psi_H(z_m), \quad \psi_B'(z_m) = \psi_H'(z_m),
\end{equation}
give the following relations; here, $(D, \lambda)$ could be either $(D_+, \lambda_+)$ or $(D_-, \lambda_-)$,
\begin{eqnarray}
\mu - \frac{\rho}{r_H^{d-2}} z_m^{d-2} &=& - \tilde z_m a - \frac{\tilde z_m^2}{2d} \left[ d (d - 3) - {\cal F}(b) \right] a, \label{mc1}
\\
- (d - 2) \frac{\rho}{r_H^{d-2}} z_m^{d-3} &=& a - \frac{\tilde z_m}{d} \left[ d (d - 3) - {\cal F}(b) \right] a, \label{mc2}
\\
D z_m^\lambda &=& b + \frac{m^2 \tilde z_m}{d} b + \frac{\tilde z_m^2}{4 d^2} \left[ m^2 (m^2 + 2d) b - \frac{a^2}{2 r_H^2} {\cal F}'(b) \right], \label{mc3}
\\
\lambda D z_m^{\lambda - 1} &=& - \frac{m^2}{d} b - \frac{\tilde z_m}{2 d^2} \left[ m^2 (m^2 + 2d) b - \frac{a^2}{2 r_H^2} {\cal F}'(b) \right], \label{mc4}
\end{eqnarray}
where $\tilde z_m = 1 - z_m$ is introduced to simplify the expressions.

By suitably combining equations~(\ref{mc3}) and~(\ref{mc4}), we can solve $D$ as
\begin{equation}
D = \frac{m^2 \tilde z_m + 2d}{d z_m^{\lambda - 1} (\lambda \tilde z_m + 2 z_m)} b,
\end{equation}
and also to get
\begin{equation} \label{a2}
a^2 = \Delta \, \frac{2 r_H^2 b}{{\cal F}'(b)},
\end{equation}
where
\begin{equation}
\Delta = \Delta(d, m, z_m) = m^2 (m^2 + 2d) + \frac{4d [ d \lambda + m^2 (\lambda \tilde z_m + z_m)]}{\tilde z_m (\lambda \tilde z_m + 2 z_m)}.
\end{equation}
The solution of $D$ shows that the critical exponent of the second-order phase transition is indeed determined by the expression of $b$. Finally, the equation for $b$ can be obtained by eliminating $a$ from eqs.~(\ref{mc2}) and~({\ref{a2}) (taking the minus square root to ensure the second-order phase transition)
\begin{equation} \label{EQb}
\sqrt{\frac{b}{{\cal F}'(b)}} \left( 1 - (d-3)\tilde z_m + \frac{\tilde z_m}{d} {\cal F}(b) \right) = \frac{(d-2) z_m^{d-3}}{\sqrt{2 \Delta}} \frac{\rho}{r_H^{d-1}}.
\end{equation}
The solution of $b$ is strongly dependent on the choice of the function ${\cal F}(\psi)$. In the next section, we will explore a general class of interesting models in detail.

\section{Explicit models}
In this section, we will explore all the possible second-order phase transitions for the models that ${\cal F}(\psi)$ is a polynomial in $\psi$.

\subsection{${\cal F}(\psi) = \psi^n$}
For the simplest case ${\cal F}(\psi) = \psi^n$, the master equation~(\ref{EQb}) for $b$ reduces to
\begin{equation}
\frac{b^{1-n/2}}{\sqrt{n}} \left( 1 - (d-3)\tilde z_m + \frac{\tilde z_m}{d} b^n \right) = \frac{(d-2) z_m^{d-3}}{\sqrt{2 \Delta}} \frac{\rho}{r_H^{d-1}}.
\end{equation}
From this equation, it is obvious that $n = 2$ is a very special value in which the overall factor $b^{1-n/2}$ disappears. Indeed, this is the only case that the second-order phase transition can happen. The value of $b$ should tend to zero near the critical point of the second-order phase transition. Therefore, as the temperature approaches to the critical value, the overall factor $b^{1-n/2}$ vanishes for $n < 2$. The above equation then implies $\rho \to 0$, which indicates that no condensates occur. For the other case $n > 2$, however, we conclude $\rho \to \infty$ near the critical point. This result seems to indicate that a condensate does appear but it is not a second-order phase transition. Naturally, one might expect that the phase transition in this case is first order. This expectation is supported by numerical results analyzed in~\cite{Franco:2009yz}. Nevertheless, our approach considered in this paper is not capable for studying these two cases.

The special case $n = 2$ indeed is just the abelian Higgs model~\cite{Franco:2009yz, Franco:2009if}, which is well studied in the literature, in particular, for dimensions $d = 3$ and $4$. There are nice reviews on this model~\cite{Hartnoll:2009sz, Herzog:2009xv, Horowitz:2010gk}. Generally, the solution for $b$ can be expressed in terms of the black hole temperature and critical temperature, $T_c$, as
\begin{equation}
b^2 = \frac{d}{\tilde z_m} \left[ 1 - (d - 3)\tilde z_m \right] \frac{T_c^{d-1}}{T^{d-1}} \left( 1 - \frac{T^{d-1}}{T_c^{d-1}} \right),
\end{equation}
where the critical temperature is given by
\begin{equation} \label{Tc}
T_c = \frac{d}{4\pi} \left[ \frac1{\sqrt\Delta} \frac{(d-2) z_m^{d-3}}{1 - (d-3)\tilde z_m} \right]^{\frac1{d-1}} \rho^{\frac1{d-1}}.
\end{equation}
It is clear that the numerical coefficients in the expressions of $b$ and $T_c$ depend on the choice of $z_m$. This reveals one weakness of this simple approach: It can only give qualitative values for the numerical coefficients. A simple comparison for $d = 3$ is given in Table I. However, our approach can capture the universal properties, in particular, the relation between the critical temperature and charge density, and the critical exponent as
\begin{equation}
T_c \sim \rho^{\frac1{d-1}}, \qquad D \sim \frac{T_c^\frac{d-1}2}{T^\frac{d-1}2} \left( 1 - \frac{T^{d-1}}{T_c^{d-1}} \right)^{1/2}.
\end{equation}

\begin{table}[ht]
\begin{tabular}{|c|c||c|c|c||l|}
\hline
\multicolumn{2}{|c|}{} & $z_m=3/4$ & $z_m=1/2$ & $z_m=1/4$ & Numerical result \\
\hline \hline
$\lambda_+ = 2$ & $\quad T_c \quad$ & 0.081 & 0.104 & 0.128 & $T_c \simeq 0.118 \sqrt{\rho}$ \\
& $\langle {\cal O} \rangle$ & 74.29 & 71.63 & 105.28 & $\langle {\cal O} \rangle \simeq 144 T_c^2 (1 - T^2/T_c^2)^{1/2}$ \\
\hline \hline
$\lambda_- = 1$ & $T_c$ & 0.114 & 0.142 & 0.161 & $T_c \simeq 0.226 \sqrt{\rho}$ \\
& $\langle {\cal O} \rangle$ & 15.20 & 11.40 & 10.05 & $\langle {\cal O} \rangle \simeq 9.3 T_c (1 - T^2/T_c^2)^{1/2}$ \\
\hline
\end{tabular}
\caption{``Qualitative'' numbers derived by the simple analytic approach for the coefficients in the critical temperature and in the condensate of the abelian Higgs model with parameters: $n = 2, d = 3, m^2 = -2$.}
\end{table}

\subsection{${\cal F}(\psi) = \psi^2 + C_\alpha \psi^\alpha$}
In this subsection, we will consider an interesting generalization to the abelian Higgs model in which the second-order phase transition can still occur. In this model, we consider ${\cal F}(\psi) = \psi^2 + C_\alpha \psi^\alpha$. The $\psi^2$ term is needed to ensure the occurrence of the second-order phase transition, and the additional term $C_\alpha \psi^\alpha$ with $\alpha > 2$ is a likely generalization. For such a model, the Eq.~(\ref{EQb}) becomes quite complicated and it is impossible to obtain an explicit solution. However, our focus is on the properties near the critical point of the second-order phase transition, where the parameter $b$ is expected to be small. Thus, Eq.~(\ref{EQb}) can be expanded in the order of the small parameter $b$. By keeping up to the leading order of $\sqrt{b/{\cal F}'(b)}$, we have
\begin{equation} \label{EQb1}
\frac1{\sqrt2} \left( 1 - \frac{\alpha}{4} C_\alpha b^{\alpha-2} \right) \left( 1 - (d-3)\tilde z_m + \frac{\tilde z_m}{d} b^2 \right) = \frac{(d-2) z_m^{d-3}}{\sqrt{2 \Delta}} \frac{\rho}{r_H^{d-1}}.
\end{equation}
It is clear that the dominant term on the left-hand side of the equation is different for the two distinct ranges of $\alpha$: (i) $\alpha > 4$ and (ii) $2 < \alpha < 4$. In addition, the particular value $\alpha = 4$ is a special case. Let us analyze these cases separately.

For the value of $\alpha$ in the range of $\alpha > 4$, the $b^{\alpha - 2}$ term is of higher order and can be neglected compared with $b^2$. Therefore, the critical exponent is exactly identical to the abelian Higgs case, namely, the mean field value $1/2$. Therefore, the additional term $C_\alpha \psi^\alpha$ with $\alpha > 4$ completely does not change any of consequences on the second-order phase transition.

For the case $2 < \alpha < 4$, instead of $b^2$, the term $b^{\alpha - 2}$ becomes dominant and Eq.~(\ref{EQb1}) reduces to
\begin{equation}
\left( 1 - \frac{\alpha}{4} C_\alpha b^{\alpha-2} \right) = \frac1{\sqrt{\Delta}} \frac{(d-2) z_m^{d-3}}{1 - (d-3)\tilde z_m} \frac{\rho}{r_H^{d-1}}.
\end{equation}
This equation allows a second-order phase transition type of solution for $b$, namely,
\begin{equation}
b^{\alpha - 2} = - \frac{4}{\alpha C_\alpha} \frac{T_c^{d-1}}{T^{d-1}} \left( 1 - \frac{T^{d-1}}{T_c^{d-1}} \right),
\end{equation}
with the same critical temperature given in~({\ref{Tc}).
To ensure that $b$ has a real value, some possible constraints on the parameter $C_\alpha$ are required; for example, $C_4$ should be negative for $\alpha = 4$. Therefore, the critical exponent in
\begin{equation}
D \sim \frac{T_c^\frac{d-1}{\alpha-2}}{T^\frac{d-1}2} \left( 1 - \frac{T^{d-1}}{T_c^{d-1}} \right)^\frac{1}{\alpha-2},
\end{equation}
is $1/(\alpha - 2)$, which is always bigger than the mean field result. Note that the exponent is discontinuous at $\alpha = 2$ of the mean field result.

Finally, $\alpha = 4$ is a special value and in such a case, we should solve
\begin{equation}
\frac1{\sqrt2} \left[ 1 - (d-3)\tilde z_m + \left( \frac{\tilde z_m}{d} - [1 - (d-3)\tilde z_m] C_4 \right) b^2 \right] = \frac{(d-2) z_m^{d-3}}{\sqrt{2 \Delta}} \frac{\rho}{r_H^{d-1}}.
\end{equation}
The coefficient of $b^2$ should be positive, which introduces a constraint to the parameter $C_4$ as
\begin{equation}
C_4 < \frac{\tilde z_m}{d [1 - (d-3)\tilde z_m]}.
\end{equation}
However, the explicit value of this constraint is not exact. It depends on the choice of $z_m$, so our analysis can only give a qualitative result. Obviously, the critical exponent is identical to the mean field result, but the numerical factor is modified
\begin{equation}
b^2 = \frac{d [ 1 - (d - 3) \tilde z_m ]}{\tilde z_m - d [1 - (d-3)\tilde z_m] C_4} \frac{T_c^{d-1}}{T^{d-1}} \left( 1 - \frac{T^{d-1}}{T_c^{d-1}} \right),
\end{equation}
which again cannot be exactly obtained by our approach.

\subsection{${\cal F}(\psi) = \psi^2 + \sum C_{\alpha_i} \psi^{\alpha_i}$}
The most general model for a polynomial ${\cal F}(\psi)$ is ${\cal F}(\psi) = \psi^2 + \sum C_{\alpha_i} \psi^{\alpha_i}$ with $\alpha_j > \alpha_i > 2$ when $j > i$. Again, the term $\psi^2$ is necessary for the existence of the second-order phase transition. It is not difficult to recognize a general feature: the much higher order extended terms are indeed more insignificant at the critical point. Therefore, for the study of universal properties, in addition to $\psi^2$, one only needs to keep the lowest order correction term, assuming $C_{\alpha_1} \psi^{\alpha_1}$. Hence, a complete analysis was already given in the previous subsection.

\section{Conclusions}
In this paper, we analytically study the universal properties of the second-order phase transition for a general class of holographic superconductors, the St\"uckelberg model with a polynomial ${\cal F}(\psi)$. The desired characteristics can be revealed by a simple approach, namely, by matching the asymptotic and horizon solutions of two essential scalar fields at an arbitrary point in between. The validity of this approach can be understood owing to the presence of a scale invariance at the critical point of the second-order phase transition. Firstly, we can classify all the possible cases admitting the second-order phase transition. Moreover, we can derive the explicit expression of the critical exponent. For most cases, the second-order phase transition has mean field critical exponent $1/2$. The essential exceptional example is the case ${\cal F}(\psi) = \psi^2 + C_\alpha \psi^\alpha$ with $2 < \alpha < 4$. In this case, the critical exponent is $(\alpha - 2)^{-1}$, which is always greater than the mean field value. All our results are consistent with the numerical analysis~\cite{Franco:2009yz, Franco:2009if}.

A significant weakness of our simple approach is that it cannot give very exact numerical coefficients since those values are dependent on the choice of the match point. Another type of analytic approach has been proposed, requiring an eigenvalue minimization while introducing a trial function~\cite{Siopsis:2010uq}. It has been checked that this approach can give very accurate results for the numerical coefficient in many models, including an external magnetic field~\cite{Ge:2010aa} and the Gauss-Bonnet gravity~\cite{Zeng:2010zn, Li:2011xj}. One should expect to improve the precision of the numerical coefficients in our results by considering other analytic approaches.

\section*{Acknowledgement}
CMC is grateful to Rong-Gen Cai and Jiro Soda for their valuable discussions and suggestions. This work was supported by the National Science Council of the R.O.C. under the grant NSC 99-2112-M-008-005-MY3 and in part by the National Center of Theoretical Sciences (NCTS).

%

\end{document}